\documentclass[aps,noshowpacs,preprint]{revtex4}

\usepackage{amsfonts}
\usepackage{amsmath}
\usepackage{amssymb}
\usepackage{graphicx}
\usepackage{bm}
\usepackage{color}

\begin{document}

\title{Nanoscale Seebeck effect at hot metal nanostructures}
\author{Aboubakry Ly$^{1}$, Arghya Majee$^{2,3}$, Alois W\"{u}rger$^{1}$}
\affiliation{$^{1}$Laboratoire Ondes et Mati\`{e}re d'Aquitaine, Universit\'{e} de
Bordeaux \& CNRS, 33405 Talence, France}
\affiliation{$^{2}$Max-Planck-Institut f\"{u}r Intelligente Systeme, 70569 Stuttgart,
Germany}
\affiliation{$^{3}$IV. Institut f\"{u}r Theoretische Physik, Universit\"{a}t Stuttgart,
70569 Stuttgart, Germany}

\begin{abstract}
We theoretically study the electrolyte Seebeck effect in the vicinity of a heated metal
nanostructure, such as the cap of an active Janus colloid in an
electrolyte, or gold- coated interfaces in optofluidic devices. 
The thermocharge accumulated at the surface varies with the local temperature, thus
modulating the diffuse part of the electric double layer. On a conducting surface 
with non-uniform temperature, the isopotential condition imposes a significant 
polarization charge within the metal. Surprisingly, this does not affect the slip velocity, 
which takes the same value on insulating and conducting surfaces. 
Our results for specific-ion effects agree qualitatively with recent observations for Janus
colloids in different electrolyte solutions. Comparing the thermal,
hydrodynamic, and ion diffusion time scales, we expect a rich transient
behavior at the onset of thermally powered swimming, extending to microseconds 
after switching on the heating.
\end{abstract}

\maketitle

\section{Introduction}

Laser-illuminated metal nanostructures provide versatile local heat engines \cite{Baf13}, with optofluidic 
applications such as trapping of nanoobjects \cite{Bra13,Lin17}, manipulation of biological cells \cite{Lin17b}, 
microflows in capillaries \cite{Bre16}, and colloidal assembly \cite{Lin17a}. Similarly, thermally powered 
artificial microswimmers rely on the conversion of absorbed heat to motion; experimental realisations include 
metal-capped Janus particles that are driven by surface forces \cite{Jia10,But12,Bar13,Sim16}, and interface 
floaters that are advected by their self-generated Marangoni flow \cite{Gir16,Wue14}.
Force-free localization and steering have been achieved by temporal \cite{Bre14}\ or spatial  \cite{Loz16} 
modulation of the laser power.

These experiments also revealed strong dependencies on material properties: Thus 
a reversal of the swimming direction was observed upon rendering the particle's 
active cap hydrophilic instead of hydrophobic \cite{But12}, or upon adding a non-ionic
surfactant to the solvent \cite{Jia10}. Similarly, copolymer coating of a
glass surface increased the thermo-osmotic velocity by one order of
magnitude \cite{Bre16}. 

Most of the cited experiments give evidence for creep flow induced by a temperature gradient in the electric 
double layer at the active surface. Very recently, a specific-ion effect was reported for silica colloids 
carrying a gold cap: their swimming velocity in a 10 mM NaCl solution changed significantly when replacing 
the cation with Lithium, or the anion with hydroxide \cite{Sim16}. These findings indicate that self-propulsion 
depends on the electrolyte Seebeck field  \cite{Wue10}, confirming previous observations on passive particles 
in an external temperature gradient, which migrated to the cold in an NaCl solution and to the hot in 
NaOH \cite{Put05,Vig10,Esl14}. Recently an enhanced Seebeck-induced flow was predicted in confined geometries \cite{Die16}. 

\begin{figure}[b]
\begin{center}
\includegraphics[width=  9cm]{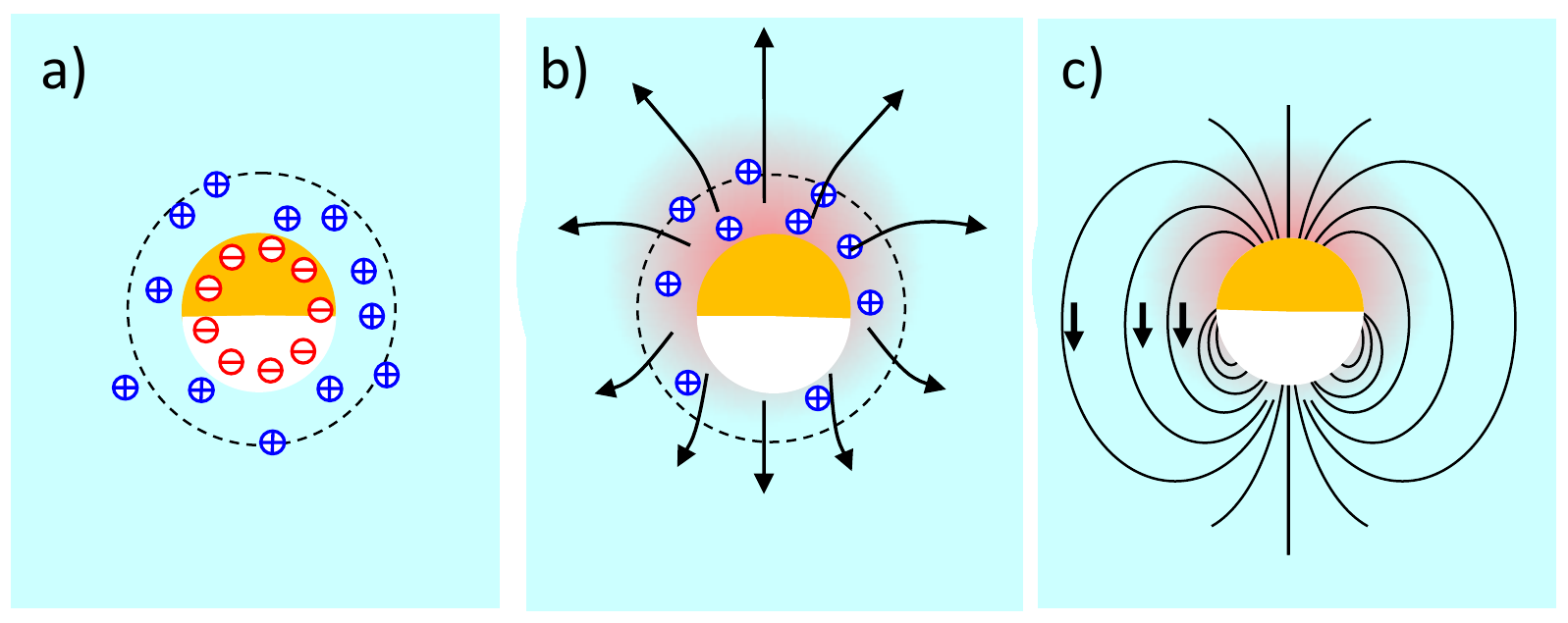}
\end{center}
\caption{Janus particle with a gold-coated upper hemisphere. 
a) The electric double layer of a micron-size particle; the diffuse layer of thickness $\lambda\sim$ a few nm, contains 
a charge $Q\sim 10^5e$. b) Upon heating the gold cap, the electrolyte Seebeck effect induces a thermocharge density 
$\rho_T$ which adds to the diffuse layer. We show the case $\rho_T>0$; for an excess temperature of a few Kelvin, 
the total thermocharge is $Q_T\sim 100e$ \protect\cite{Maj12}. The corresponding negative ions are at the boundary 
of the experimental cell. The arrows indicate the thermoelectric field. c) Schematic view of the thermoelectric 
field after subtraction of the monopole term $\propto Q_T/r^2$.  The diffuse layer is not shown. The parallel 
component $E_\parallel$ vanishes at the conducting gold surface; at larger distance one has the dipolar field $\propto r^{-3}$.}
\end{figure}

In this paper we study how the electrolyte Seebeck effect modifies the electric 
double layer and drives a creep flow along a surface with non-uniform temperature. 
The main features are illustrated in Fig. 1 at the example of a gold-capped Janus particle, 
but are  generally valid for metal nanostructures in contact with water \cite{Bra13,Lin17,Lin17b,Bre16,Lin17a}. 
Upon heating the gold cap with a laser, the salt ions move along the temperature gradient, 
and an excess charge $Q_T$ forms at the hot surface, as shown in the middle panel; the 
corresponding negative ions are at the wall of the container. The resulting electric 
field comprises, besides the radial monopole term $\propto Q_T/r^2$, a parallel component 
along the particle surface; the latter exerts a force on the double layer and induces creep flow. 

We address two main questions: First, how are the double layer and the Seebeck field 
modified by the electrostatic boundary conditions at insulating and conducting surfaces? 
Second, the equipotential condition at a conductor requires a zero parallel electric field, 
as illustrated for the upper hemisphere in Fig. 1c. Does this imply that the thermoelectric 
creep velocity is suppressed at the gold cap of Janus particles?

The outline of our paper is as follows. In Sect. II we briefly review the bulk electrolyte 
Seebeck effect, where boundary effects are irrelevant. In Sect. III we evaluate the thermocharge 
and the Seebeck near-field at a surface, which are sketched in Fig. 1 b and c. Starting from the 
integral expression of Gauss' law, the thermoelectric properties and the modification of the 
double-layer are derived both for insulating and conducting surfaces. Sect. V is devoted to the 
thermodynamic forces resulting from the non-equilibrium state of the double layer, and to the 
creep flow along the surface. Novel results arise from the parallel component of the thermoelectric 
and polarization fields derived in Sect. IV. In the final sections we discuss and summarize our results. 

\section{Electrolyte solution in a temperature gradient}

We briefly review the steady-state response of an electrolyte solution to a non-uniform 
temperature \cite{Gro62}, the resulting Soret and Seebeck effects, and in particular the thermoelectric field. 

\subsection{Thermodynamic forces}

Consider monovalent ions with concentrations $n_\pm$, enthalpy $H_\pm$, and chemical potential 
$\mu_\pm=H_\pm + k_BT\ln  n_\pm$. Then the ions are subject to the thermodynamic forces, which derive from 
the Planck potential $\mu_\pm/T$ \cite{Gro62},
    \begin{equation}
     -T\mathbf{\nabla} \frac{\mu_\pm}{T} =   
          - k_BT\frac{ \mathbf{\nabla} n_{\pm }}{n_\pm}    + H_{\pm }\frac{\mathbf{\nabla }T}{T} , 
   \label{eq0}
   \end{equation}
where the first term in (\ref{eq0}) accounts for gradient diffusion, and the second one for thermodiffusion 
along the temperature gradient. The prefactor of the latter arises from the Gibbs-Helmholtz equation 
$d(\mu_\pm/T)/dT=-H_\pm/T^2$. Note that this relation does not imply constant enthalpies; the quantities 
$H_{\pm }$ may depend on temperature. 

These thermodynamic forces give rise to ion currents $\mathbf{J}_{\pm }$. When including an electric 
field $\mathbf{E}$ we find 
    \begin{equation}
     \mathbf{J}_{\pm } 
                = m_{\pm } \left( - k_B T \nabla n_{\pm } + n_\pm H_\pm \frac{\mathbf{\nabla }T}{T} \pm  en_\pm \mathbf{E} \right) ,  
    \label{eq1}
    \end{equation}
where we have assumed that the mobilities $m_{\pm }$ are the same for thermodynamic and electric forces, 
and are related to the diffusion coefficients by $m_{\pm } = D_{\pm }/k_BT$. The steady state is, in general, 
characterized by the condition of constant currents with zero divergency, $\rm{\nabla}\cdot\mathbf{J}_{\pm } = 0$. 
In the case of a closed system with solid boundaries, and in the absence of external forces acting on the ions, 
however, there is no source field and the currents vanish. In this preliminary section, we consider 
non-interacting boundaries, and thus put $\mathbf{J}_{\pm } = 0$.

\subsection{Salt Soret effect}

It turns out convenient to consider the salinity $n=(n_++n_-)/2$ and the charge density 
$\rho=e(n_+-n_-)$ rather than the ion concentrations $n_\pm$. Then the sum of $\mathbf{J}_\pm=0$ 
provides the ``Soret equilibrium'' for the salinity, 
   \begin{equation}
         \nabla n + n S_T \mathbf{\nabla }T  = 0 ,
   \label{eq6} 
   \end{equation}
with the salt Soret coefficient
   \begin{equation}
   S_T= \frac{H_{+}  +  H_{-}}{2k_BT^2}.
   \label{} 
   \end{equation}
Eq. (\ref{eq6}) implies a salinity gradient throughout the sample. Since the enthalpies $H_\pm$ are 
of the order of $k_B T$, the relative salinity change is comparable to the relative excess temperature, 
$\Delta n/n\sim\Delta T/T$. Soret data for various salts were first reported by Chipman in 1926 \cite{Chi26}.

\subsection{Electrolyte Seebeck effect and surface charges}

Now we consider the difference of the equations $\mathbf{J}_\pm=0$, which result in a relation for the 
stationary charge density and electric field. Far from the boundaries, the charge density $\rho$ must 
vanish because of the huge cost in electrostatic energy required by charge separation. Then we find that, 
in order to satisfy the zero-current condition, the temperature gradient is accompanied by a constant 
bulk electric field,    
    \begin{equation}
    \mathbf{E}_T = S\mathbf{\nabla }T,
   \label{eq4}
   \end{equation}
with the coefficient 
   \begin{equation}
    S = - \frac{H_{+} - H_{-}}{2eT}.
   \end{equation} 
$\mathbf{E}_T$ is called the macroscopic thermoelectric field, in analogy to the Seebeck effect in metals 
and semiconductors \cite{Put05}. In the latter, the Seebeck coefficient is determined by the temperature 
dependence of electronic properties, whereas for an electrolyte solution, $S$ is given by the difference 
of ion enthalpies. Depending on the $H_\pm$, the Seebeck coefficient may take either sign; typical values 
are of the order of  $10^{-4}\,\mathrm{V/K}$ \cite{Wue10}. In the literature one often finds the ``heat of 
transport'' $Q_\pm = - H_\pm$ with the opposite sign; the most complete data so far are reported in 
Ref.~\cite{Nak88}. The above derivation of the Seebeck field has first been given by Guthrie \cite{Gut49}, 
relying on the conditions of zero ion currents and zero charge.  

Like any static electric field, $\mathbf{E}_T$ must originate from positive and negative charges. Starting 
from $\mathbf{J}_\pm=0$ and allowing for finite $\rho$, we obtain a relation for the stationary charge 
density and electric field,
   \begin{equation}
         \nabla \rho + \frac{\varepsilon}{\lambda^2}  ( S\mathbf{\nabla }T - \mathbf{E} )= 0 ,  
   \label{eq2}
   \end{equation}
with the Debye length $\lambda^2=\varepsilon k_BT/2ne^2$. Adding Gauss' law 
   \begin{equation}
   \mathbf{\nabla}\cdot\varepsilon\mathbf{E}=\rho
   \label{eq3}
   \end{equation}
one finds that the only solution in the bulk corresponds to  (\ref{eq4}) with $\rho=0$. At the hot and cold 
boundaries, however, there are finite thermocharge densities $\rho_T$ of opposite sign. In physical terms, 
the thermocharges originate from the unlike thermodiffusion of the cations and anions in (\ref{eq1}).

We briefly summarize the above derivation of the electrolyte Seebeck effect. It arises from the tendency of 
salt ions to migrate along a temperature gradient. The underlying thermodynamic forces $H_\pm \nabla T/T$ 
follow from the entropy balance of the non-equilibrium electrolyte solution \cite{Gro62}.  Regarding the 
salt concentration $n=\frac{1}{2}(n_++n_-)$, the Soret equilibrium (\ref{eq6}) describes the stationary 
salinity gradient; in physical terms it satisfies the steady-state condition, requesting that diffusion and 
thermodiffusion currents of salt cancel each other.  

The Seebeck effect presents a more intricate situation, since it stems from the difference of cation and anion 
currents. An enthalpy difference $H_+\ne H_-$, tends to partly separate positive and negative ions. As an 
important consequence, this results in surface charges and a macroscopic thermoelectric field. Thus one has 
to satisfy Gauss' law, in addition to the steady-state condition. 

For a negative Seebeck coefficient, the thermodiffusion currents result in positive  and negative charges at 
the hot and cold boundaries, respectively. In the case of a heated particle in a bulk electrolyte solution, 
the hot boundary reduces to the particle surface, which accordingly is covered by a diffuse layer of mobile 
cations, as illustrated in Fig. 1b. Then the particle carries a net thermocharge which is related by Gauss' 
law to a monopole  field that decays  as $r^{-2}$ with the distance $r$ \cite{Maj12}; the field lines end 
at the corresponding anions which are at the wall of the experimental cell. In the present paper we are 
concerned with the dipolar contribution of the Seebeck field, which is sketched in Fig.~1c. 

The linear equations (\ref{eq2}) and (\ref{eq3}) correspond to the Debye-H\"uckel approximation. Their solution is 
generally valid at otherwise uncharged boundaries. Simple 1D and radially symmetric 3D geometries have been studied 
previously in \cite{Maj11,Maj12}. The general case of an uncharged surface is treated in 
Sect. \ref{subsect:uncharged surface} and in Appendix A. A more complex situation occurs at charged surfaces, since 
the diffuse layer comprises the counterions and the thermocharge; in the following section this is treated in 
non-linear Poisson-Boltzmann theory.

\section{Thermocharge and thermoelectric near-field}

Here we evaluate how the thermoelectric properties at the particle surface depend on the material 
properties, and in particular on its surface charge and electrical conductivity. We first write the usual boundary 
layer approximation in a form that is well adapted to the condition imposed by the Seebeck far-field.

Thus we calculate the thermocharge density $\rho_T$ and the thermoelectric field in the vicinity of the surface. In 
order to clearly separate the charge effects induced  by the temperature gradient from those of the electric double 
layer, we first study an insulating particle that does not carry surface charges. The strong permittivity contrast 
between water and typical materials such as polystyrene or silica, simplifies the electrostatic boundary conditions. 

Then we consider charged surfaces and, moreover, distinguish insulating and conducting materials. The main difficulty 
arises from the fact that the diffuse layer contains both the counterions of Fig. 1a and the thermocharge of Fig. 1b, 
which have to be treated on an equal footing in terms of Poisson-Boltzmann theory.

\subsection{Boundary layer approximation}

Surface charges of colloidal particles are screened by a diffuse layer of counterions. An analytic mean-field 
solution exists in one dimension only. It provides a controlled approximation at curved surfaces, as long as the 
local curvature radius is much larger than the Debye screening length $\lambda$. Then there is a separation of 
length scales: The properties of the electric double layer vary much more rapidly in perpendicular direction than 
parallel to the surface. 

The resulting approximation is best discussed in terms of Gauss' law (\ref{eq3}). The normal field component 
varies on the scale of $\lambda$, whereas the permittivity and the parallel electric field vary on the scale 
of the particle radius $a$. Thus to linear order in $\lambda/a$, Gauss' law simplifies to 
  \begin{equation}
   \frac{dE_{\perp}}{dz} = \frac{\rho}{\varepsilon} ,
  \end{equation}
where $z$ is the distance from the surface. Here and in the following, $E_{\perp}$ points away from the surface; thus 
for a spherical particle,  $E_{\perp}$ is the radial component, and $z=r-a$. 
  
For further use, we integrate from the surface to a distance $B$ that is much larger than the 
screening length but much smaller than the particle radius, $\lambda\ll B\ll a$, and find
  \begin{equation}
  {E}_{\perp}(B) - {E}_{\perp}(0) =  \frac{1}{\varepsilon}\int_0^B dz \rho(z) 
                                                             \equiv \frac{\sigma}{\varepsilon} .
  \label{eq7}
  \end{equation}
The second identity defines the charge density per unit area of the diffuse layer. This parameter 
also determines the double-layer potential $\varphi_\sigma$, as is obvious from the  
Poisson-Boltzmann mean-field expression (\ref{B2}) for the diffuse layer.

In the case of an electric double layer at equilibrium, the electric field vanishes at large distance, 
${E}_{\perp}(B)=0$, resulting  at the particle surface in ${E}_{\perp}(0) =  - \sigma/\varepsilon$.
Then $-\sigma$ corresponds to the charge per unit area of the surface, which exactly cancels 
that of the diffuse layer. 

On the contrary, the main results of the present paper are derived from Eq. (\ref{eq7}), 
with the outer boundary condition determined by the thermoelectric far-field (\ref{eq4}). 
This implies that $\sigma$ as defined in (\ref{eq7}) contains counterions and thermocharge, and thus 
does no longer define the surface charge density.

\subsection{Uncharged insulating surface}\label{subsect:uncharged surface}

Because of the strong permittivity contrast of water and silica or polystyrene, the Seebeck field hardly 
penetrates the surface. Then the electrostatic boundary conditions require that the normal electric field 
vanishes at the surface, whereas at the outer boundary one has the bulk Seebeck field, 
   \begin{equation}
   E_{\perp}(0)=0, \;\;\;\;   E_{\perp}(B) =  S\nabla_\perp T_S. 
  \end{equation}
In the outer boundary condition we have used that the temperature gradient at $B$ (with $B\ll a$) hardly differs 
from its value at the surface. In other words, the temperature gradient $\nabla_\perp T$ may be taken as constant 
well beyond the charged layer. 

From Gauss' law (\ref{eq7}) one readily finds        
  \begin{equation} 
  \varepsilon S\nabla_\perp T_S=  \int_0^B dz \rho_T(z) \equiv \sigma_T ,
  \label{eq8} 
  \end{equation}
where the second equality defines the thermocharge per unit area. Since the temperature decreases with the 
distance from the surface, the outward component of the gradient is negative,  $\nabla_\perp T<0$. Thus a 
negative Seebeck coefficient implies a positive surface charge at the hot boundary, $\sigma_T>0$, as illustrated in Fig. 1b. 

In general, the temperature varies also along the particle surface, and so does $\sigma_T$, as illustrated in Fig. 1b. 
As a consequence, the Seebeck field  is not radially symmetric. In particular, the difference in thermocharge between 
the upper and lower hemispheres is at the origin of the dipolar field component shown in Fig.~1c.  

In physical terms,  the thermocharge screens the Seebeck field as one 
approaches the solid boundary. For a micron size particle at an excess temperature of 10 K, and  a typical Seebeck parameter 
$S=10^{-4}\,\mathrm{V/K}$, the surface charge density $\sigma_T$ takes a value of about $10e$ per 
square micron and the electric field about 1 kV/m. Because of its small value, the thermocharge
is well described by Debye-H\"uckel theory with an exponential decay,
  \begin{equation}
  \rho_T(z)= e^{-z/\lambda}\sigma_T/\lambda. 
  \end{equation}
One readily finds that the normal component of the electric field  is screened by the thermocharge
such that it vanishes at the surface
   \begin{equation}
    E_\perp^T =S\nabla_\perp T (1 - e^{-z/\lambda}). 
  \label{eq9} 
\end{equation}
The parallel component, on the other hand, remains unchanged and is finite at the surface,
  \begin{equation}
    E_\parallel ^T =S\nabla_\parallel  T_S.
   \label{eq9a} 
\end{equation}

These equations express thermocharge and Seebeck field through local quantities. In Appendix A we rederive these 
quantities in terms of a multipole expansion for a spherical particle.  The above $E_\perp^T$ has been obtained 
previously \cite{Maj11,Maj12} for simple geometries where $E_\parallel ^T =0$ .

\subsection{Charged insulating surface}

Now we consider an insulating surface with an electric double layer. We assume a negative surface charge 
density $-\sigma_0$,  as is the case for most colloids. Then the electric field satisfies the boundary conditions
   \begin{equation}
   E_{\perp}(0)=-\sigma_0/\varepsilon, \;\;\;\;   E_{\perp}(B) =  S\nabla_\perp T_S. 
  \end{equation}
From Gauss' law (\ref{eq7}) one readily finds
  \begin{equation}
  S\nabla_\perp T_S + \frac{\sigma_0}{\varepsilon} 
                           =  \frac{1}{\varepsilon}\int_0^B dz (\rho _{T}+\rho _{0}) 
                          =   \frac{\sigma_I}{\varepsilon} ,
  \label{eq30}
  \end{equation}
with the charge density of mobile ions per unit area,
  \begin{equation}
  \sigma_I=\sigma_T+\sigma_0,
  \label{eq31}
  \end{equation}
consisting of the thermocharge and the particle's counterions. 

The corresponding Poisson-Boltzmann potential $\varphi_{\sigma_I}$, which is defined through 
$\rho _{T}+\rho _{0}=-\varepsilon \partial_z^2\varphi_{\sigma_I}$, has to be calculated with an effective 
parameter ${\sigma_I}$, which is different from the actual surface charge $-\sigma_0$. Then we have the total potential 
  \begin{equation}
  \varphi_I = \varphi_T+ \varphi_{\sigma_I}.
  \label{eq32}
  \end{equation}
The normal component of the electric field reads
  \begin{equation}
  E_{\perp}(z) =  S\nabla _\perp T - \nabla_\perp\varphi_{\sigma_I}.
  \end{equation}
The second term decays rapidly through the screening layer, where the first one is constant on the 
scale of the Debye length. With the explicit result (\ref{B5}) for the second term, the near-field takes the simple form
\begin{equation}
   E_\perp(z)  = \frac{\sigma_T}{\varepsilon } -\frac{\sigma_0 
          + \sigma_T}{\varepsilon }e^{-z/\lambda }\frac{1-g^{2}}{1- \hat{g}^{2}},  \;\;\;\;\;  (z\ll a) ,
  \end{equation}
with $\hat{g}=e^{-z/\lambda}g$ and the parameter $g$ as defined in (\ref{B3a}). One readily verifies that $E_{\perp}$  satisfies 
the above boundary conditions. 

The parallel component of the electric field,
  \begin{equation}
  E_{\parallel}(z)= S\nabla _{\parallel }T_S - \nabla _{\parallel }\varphi_{\sigma_I} ,
    \label{eq33}
  \end{equation}
does not vanish at the surface $z=0$. The explicit form of the second term $\nabla _{\parallel }\varphi_{\sigma_I}$ could be 
readily calculated from the Poisson-Boltzmann potential (\ref{B3}); it turns out that it is small as compared to the bare Seebeck 
field,   
  \begin{equation}
  E_{\parallel}(z)= S\nabla _{\parallel }T_S (1+ O(\lambda/a)),
  \label{eq:E_ins}
  \end{equation}
and thus may be discarded.

\begin{figure}[ptb]
\begin{center}
\includegraphics[width=9cm]{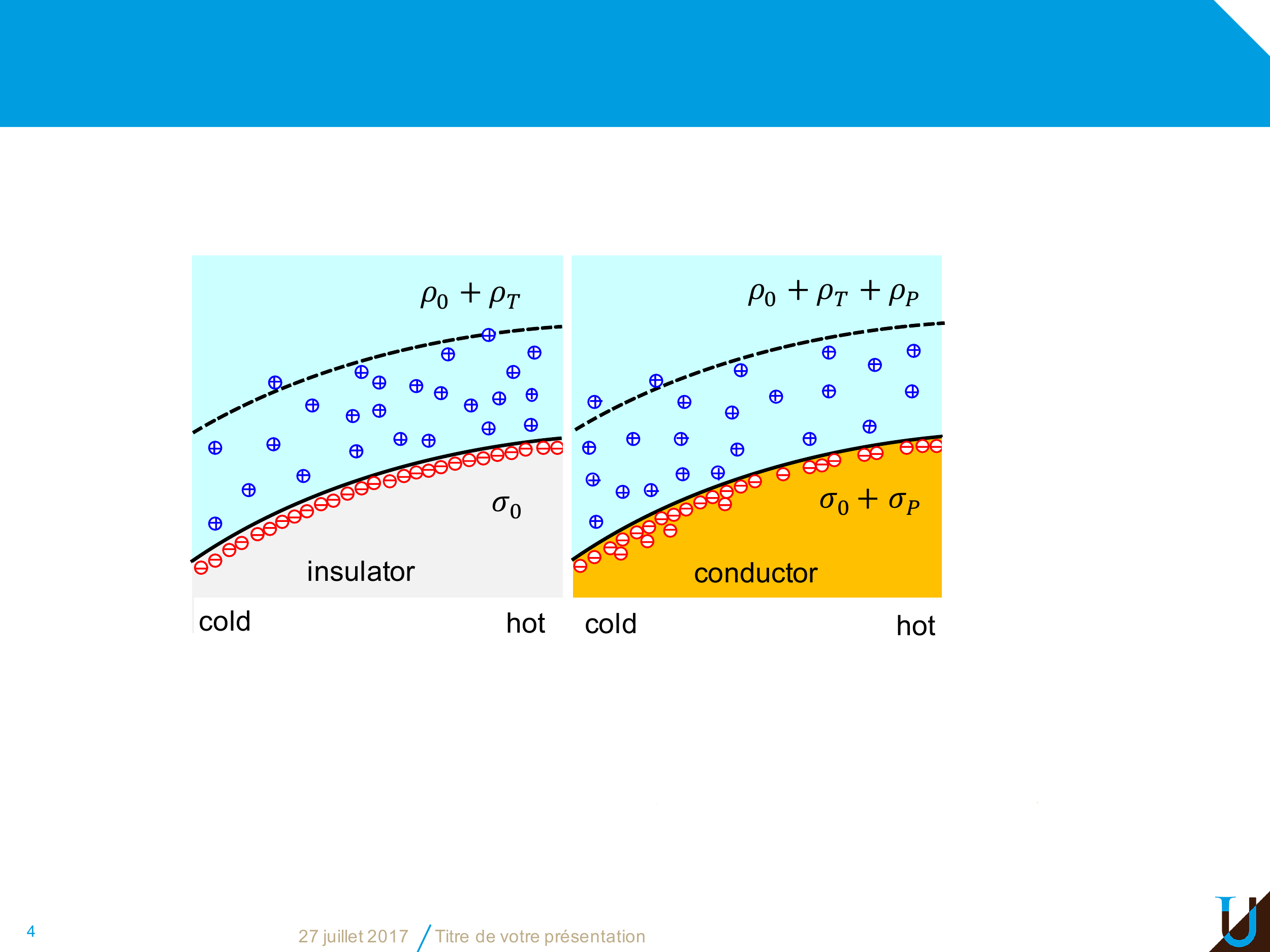}
\end{center}
\caption{Electric double-layer in a thermal gradient (temperature
increases from left to right). The left panel shows an insulating material. The surface charge
density $\protect\sigma _{0}$ is not affected by the Seebeck effect, whereas
the diffuse layer comprises the non-uniform thermocharge density $\protect\rho _{T}$. 
Its absolute value $|\rho_T|$ is proportional to the excess surface temperature, and its sign depends on 
the Seebeck coefficient; we show the case $S<0$. At a conducting surface (right panel), the
parallel component of the electric field vanishes. The condition (\ref{6}) requires a polarization charge 
$\protect\sigma_{P}$ which induces a corresponding 
displacement of counterions $\rho_{P}$.  Thus the diffuse layer consists of the counterions of 
$\sigma_0$ and $\sigma_P$,  and of the thermocharge $\rho_T$. For typical parameters, these 
contributions satisfy $|\rho_T|\ll |\rho_P| \ll |\rho_0|$. }
\end{figure}

\subsection{Charged conducting  surface}

Now we turn to conducting surfaces, such as the gold cap of the upper hemisphere in Fig. 1c.
The electrostatic boundary conditions impose a constant potential, or a vanishing 
parallel electric field  \cite{And06}, whereas at the outer boundary  $z=B$, it is given by the 
Seebeck far-field:
  \begin{equation}
  E_\parallel (0) = S\nabla_\parallel T_S - \nabla_\parallel \varphi_{\sigma_C}(0) = 0,
                                          \;\;\;\;\;  \mathbf{E} (B) = S\mathbf{\nabla} T. 
  \label{6}
  \end{equation}
These conditions cannot be satisfied with the constant surface charge $-\sigma_0$ discussed so far. 

To achieve (\ref{6}) the mobile electrons in the metal surface move until their polarization
charge density $\sigma _{P}$ results in a constant surface potential. The polarization charge is determined by inserting 
$\varphi_\sigma$ with 
  \begin{equation}
  {\sigma_C}(x) = \sigma_T(x)+\sigma_0+\sigma_P(x)
  \label{eq34}
  \end{equation} 
in Eq. (\ref{6}) and solving for $\sigma_P$. Assuming that the total charge does not change, one 
has for the surface integral $\left< \sigma_P \right>=S^{-1}\int dS\sigma_P=0$. Its derivation is given in Appendix C. 
Its overall behavior is illustrated by the simpler expression (\ref{eq92}) obtained in Debye-H\"uckel approximation,
  \begin{equation}
   \sigma_P = \frac{\varepsilon S(T_S - \left< T_S \right> )}{\lambda} ,  \;\;\;\;\;\; (\mathrm{DHA}).
   \label{eq35}
  \end{equation}
The polarization charge varies along the surface and even changes sign. For a negative Seebeck coefficient, 
one has $\sigma_P<0$ at the hot end of the metal surface, and $\sigma_P>0$ at the cold end, as shown in Fig.~2b. 

Since the diffuse layer screens the local surface charge density, $\sigma_{P}$ induces a corresponding 
change of the mobile charge density, $\rho_{P} $, and we have $\rho_C  = \rho _T+\rho _{0}+\rho _P$.  
We recall that the double-layer potential $\varphi_\sigma$ is calculated with the parameter $\sigma_C$ 
which accounts for the charge density of the diffuse layer, $\sigma_C=\int dz \rho_C(z)$, whereas the 
surface charge density is given by $-(\sigma_0+\sigma_P)$. Accordingly, we have 
  \begin{equation}
  E_{\bot }(0) =- \frac{\sigma_{0}+\sigma_{P}}{\varepsilon}
  \label{26}
  \end{equation}
at the particle surface.

The parallel field component of the electric field,
  \begin{equation}
  E_\parallel(z)  
                =  S\mathbf{\nabla}_\parallel T_S - \mathbf{\nabla}_\parallel \varphi_{\sigma_C}(z),
  \label{27}
  \end{equation}  
is zero at the particle surface. With increasing distance, the double-layer potential $\varphi_\sigma$ 
decays and vanishes well beyond the screening length, and the electric field is given by (\ref{eq4}). The overall 
behavior is best displayed  in Debye-H\"uckel approximation, 
  \begin{equation}
  E_\parallel(z)  
                =  S\mathbf{\nabla}_\parallel T_S \left( 1  - e^{-z/\lambda}\right) ,\;\;\;\; (\mathrm{DHA}).
  \label{eq:E_cond}
  \end{equation}  
This expression satisfies both the surface and far-field boundary conditions (\ref{6}). The crossover occurs at 
the scale of the Debye length and results from the polarization charge $\sigma_P$, whereas the far-field is related 
to the thermocharge $\sigma_T$.

\section{Non-equilibrium double-layer and creep flow}

In the absence of interactions between the electrolyte solution and the boundaries, the stationary state is 
characterized by a salt gradient, a Seebeck field, and thermocharges at the boundaries, but there is no flow 
or electric current; compare the steady state obtained in Sect. II. Now we turn to interacting surfaces, more 
precisely to charged boundaries with an electric double layer, and we derive the creep flow along the surface. 
We linearize in the gradients of the non-equilibrium state; this implies that we do not consider the coupling 
of the Seebeck field with the thermocharge. 

\subsection{Thermodynamic forces and slip velocity}

Closely following Ref. \cite{Wue10}, we derive how the electric double layer of the surface interacts with the 
temperature gradient and its companion fields.  Novel results arise from the coupling of the diffuse layer with 
the Seebeck field.  We start from the well-known expression for the effective slip velocity \cite{Der41,And89},
  \begin{equation}
  v_s =  \frac{1}{\eta} \int_0^\infty dz z f_\parallel ,  
   \label{eq40}
  \end{equation}
where $\eta$ is the solvent viscosity and $f_\parallel$ the parallel component of the thermodynamic force density 
arising from the non-equilibrium state.  

The force acting on a unit volume of the electric double layer comprises the divergency of the Maxwell tensor 
$\mathcal{T}$ and the gradient of the osmotic pressure $P$, 
  \begin{equation}
  \mathbf{f} = \mathbf{\nabla}  \cdot \mathcal{T} - \mathbf{\nabla} P  .
   \label{eq41}
  \end{equation}
The former accounts for the electric energy of the double layer; the resulting force 
  \begin{equation}
  \mathbf{\nabla}  \cdot \mathcal{T} =  \rho \mathbf{E} - \frac{1}{2}E^2 \mathbf{\nabla}\varepsilon 
                                                        =  \rho (S\mathbf{\nabla} T - \mathbf{\nabla} \varphi_\sigma) 
                                                                 - \frac{1}{2}E^2 \mathbf{\nabla}\varepsilon,
   \label{force}
  \end{equation}
consists of the Coulomb force on the diffuse layer and the change in electric energy due to a permittivity 
gradient \cite{Str41,Lan87,Fay08}. The second equality separates the double-layer and Seebeck contributions 
to the Coulomb force.  

The second term in (\ref{eq41}) stems from the osmotic pressure $P=\delta nk_BT$ exerted by the excess ion 
concentration  $\delta n$ in the double layer. Inserting (\ref{B8}) and evaluating the gradient, one needs 
to account for the variation with temperature, salinity, and the potential $\varphi_\sigma$, resulting in 
  \begin{equation}
  \mathbf{\nabla}  P = -\rho \mathbf{\nabla} \varphi_\sigma  
                                              +(\rho\varphi_{\sigma} + \delta nk_BT) \frac{\mathbf{\nabla}  T}{T}
                                              + \delta n k_BT  \frac{\mathbf{\nabla}n}{n}.
  \end{equation}
In these relations for $\mathbf{\nabla}  \cdot \mathcal{T}$ and $ \mathbf{\nabla}  P$, the potential $ \varphi_\sigma$ 
varies rapidly in normal direction, and slowly along the surface. The quantities $T$, $\varepsilon$, and $n$ 
vary slowly in all directions, on the scale of the particle parameter, whereas the charge density $\rho$ and the ion 
density $\delta n$ vanish beyond the diffuse layer. 

Gathering the different terms one obtains the force density 
  \begin{equation}
    \mathbf{f}  =  \rho S \mathbf{\nabla}  T - (\rho\varphi_{\sigma} + \delta nk_BT) \frac{\mathbf{\nabla} T}{T}
                                    - \delta n k_BT  \frac{\mathbf{\nabla}n}{n}
				   - \frac{1}{2} E^2\mathbf{ \nabla} \varepsilon.
   \label{eq46}
  \end{equation}
In addition to the temperature gradient, $ \mathbf{f} $ depends on the gradients of salinity and permittivity, 
induced by the Soret effect and the temperature dependence of $\varepsilon$. In linear-response approximation, 
we replace the coefficients of the gradients in (\ref{eq46}) by the corresponding equilibrium quantities, and 
the electric field in the last term by $-\nabla\varphi_{\sigma_0}$. The gradient fields in (\ref{eq46}) are 
constant on the scale of the screening length, whereas the coefficients $\rho$, $\delta n$, and $E$ vanish well 
beyond the diffuse layer. 

As a remarkable feature, the parallel gradient $\nabla_\parallel \varphi_{\sigma}$ has  disappeared from the 
double-layer forces. While both the electrostatic force $\nabla\cdot\mathcal{T}$ and the pressure gradient $\nabla P$ 
depend on the precise form of the parameter $\sigma$, these terms cancel in (\ref{eq46}), and so do the polarization 
contributions. With the Poisson-Boltzmann expressions for $\varphi_{\sigma}$ and its derivatives given in Appendix B, 
the integrals in (\ref{eq40}) are readily performed \cite{Mor99,Wue08},
  \begin{equation}
  v_s =  -  \frac{\varepsilon\zeta} {\eta}S \nabla_\parallel  T 
              + \frac{\varepsilon(\zeta^2-3\zeta_T^2)}{2 \eta} \frac{ \nabla_\parallel  T} {T}
              - \frac{\varepsilon\zeta_T^2}{2 \eta} 
                 \left(\frac{\nabla_\parallel  \varepsilon} {\varepsilon} +\frac{\nabla_\parallel  n} {n} \right),
   \label{eq48}
  \end{equation}
with the surface potential $\zeta=\varphi_{\sigma_0}(0)$ and the quantity $\zeta_T=(2k_BT/e)[\ln\cosh(e\zeta/4k_BT)^2]^{1/2}$. 
Each term of the slip velocity consists of a gradient field characterizing the non-equilibrium state of the electrolyte solution, 
and a coefficient that depends on the equilibrium properties of the solid surface and of the electrolyte solution. 
With the bulk salinity gradient $\nabla n$ as defined in (\ref{eq6}) and the logarithmic permittivity derivative 
$\tau=-d\ln \varepsilon/d\ln T$, one has 
$$\frac{\nabla_\parallel  n}{n} = - S_T{\nabla_\parallel  T},  
                      \;\;\;\;\; \frac{\nabla_\parallel \varepsilon}{\varepsilon} = - \tau  \frac{\nabla_\parallel  T}{T}, $$
where $\tau\approx 1.5$ at room temperature \cite{Cat03}. A temperature gradient  of Kelvin per micron results in a velocity of 
micron per second. 

At a surface in a constant external temperature gradient $\nabla T$, the parallel component is simply given by its 
projection on the surface; for a spherical particle one has $\nabla_\parallel  T= \sin\theta \nabla T$, with the 
polar angle $\theta$ and where we have discarded corrections due to the thermal conductivity contrast; see Eq. (\ref{eq74}) 
below. The self-generated temperature field of a laser-heated particle results in a more complex 
expression, depending on its absorption coefficient and thermal conductivity \cite{Bic13}. The surface potential 
$\zeta$ usually depends weakly on temperature; the variation of $v_s$ is rather irrelevant except for Janus 
particles with different $\zeta$ on the two hemispheres; the surface potential could even take opposite signs 
on the metal cap and on the insulating half.  

The novel result concerns the thermoelectric contribution to (\ref{eq48}), that is, the first term proportional to 
the electrolyte Seebeck coefficient $S$. The remaining term $\propto \nabla T$ and that $\propto \nabla \varepsilon$ 
are known as thermo-osmosis \cite{Der41,Bre16}, whereas the last one, $\propto \nabla n$, is similar to salt 
osmosis \cite{Pri87,Abe09}. As a main finding of this work, we note that $v_s$ does not depend on the electrical 
conductivity of the particle surface. The slip velocity is the same for insulating and conducting materials, although 
the electric field at the surface shows quite a different behavior: Its parallel component is finite at an insulating 
surface but vanishes at a conductor, as shown by Eqs. (\ref{eq:E_ins}) and (\ref{eq:E_cond}), respectively. A similar 
effect was shown to occur for the electrophoretic mobility at a metal surface \cite{Squ04}, resulting in an electroosmotic 
slip velocity that is the same at insulating and conducting surfaces. 

\begin{figure}[ptb]
\begin{center}
\includegraphics[width=16cm]{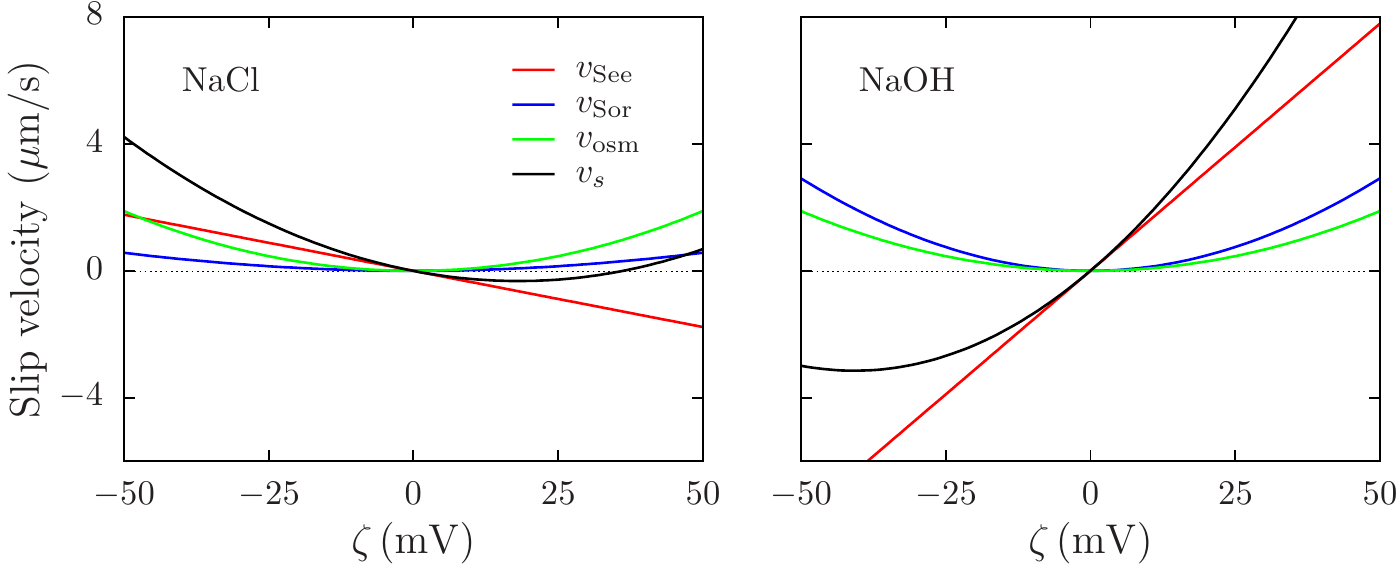}
\end{center}
\caption{Comparison of the different contribution to the slip velocity (\ref{eq49}) for NaCl and NaOH solutions, as a 
function of the $\zeta$-potential. We plot separately the Seebeck, Soret, and thermoosmotic contributions, as defined 
in (\ref{eq49}). For NaCl and NaOH solutions, the Seebeck coefficient takes the values  $S=-0.2\,\textrm{mV/K}$ and 
$S=0.05\,\textrm{mV/K}$, respectively; for the Soret coefficent one has $S_T= 2.7\times 10^{-3}\,\textrm{K}^{-1}$ and 
$1.4\times 10^{-2}\,\textrm{K}^{-1}$. We use the temperature gradient $\nabla_\parallel T=1\,\mathrm{K/\mu m}$, the 
viscosity and permittivity of water, and ambient temperature.}  
\label{Fig:vs}
\end{figure}

\subsection{Relevance of ion-specific contributions}

In order to compare their relative importance, we plot in Fig. \ref{Fig:vs} the different contributions to the slip 
velocity, for parameters describing NaCl and NaOH solutions. With a temperature gradient of 1K/$\mu$m, which is easily 
achieved by heating gold microstructures, one finds velocities of the order of microns per second. We split the slip 
velocity (\ref{eq48}) in three terms, 
\begin{equation}
     v_\mathrm{See}=  -  \frac{\varepsilon\zeta} {\eta}S \nabla_\parallel  T, \;\;\;\;  
      v_\mathrm{Sor} =  \frac{\varepsilon\zeta_T^2}{2 \eta}    S_T{\nabla_\parallel  T},\;\;\;\;  
      v_\mathrm{osm} =  \frac{\varepsilon[\zeta^2-(3-\tau)\zeta_T^2]}{2 \eta} \frac{ \nabla_\parallel  T} {T},
      \label{eq49}
\end{equation}
where the first and second ones are proportional to the Seebeck and Soret coefficients, and the third one describes 
the velocity induced by heat flow or ``thermoosmosis''. This thermoosmotic velocity $v_\mathrm{osm}$ is dominant 
in the absence of salt \cite{Der41,Bre16}. In the presence of salt, however, the Seebeck and Soret velocities exceed 
thermoosmosis; experiments on nanometric micelles \cite{Vig10} and micrometric polystyrene particles \cite{Esl14} 
provide conclusive evidence for magnitude of the ion-specific Seebeck and Soret contributions. The data of 
Ref. \cite{Esl14} indicate that both $S$ and $S_T$ strongly depend on temperature. 

Note that the Seebeck term is linear in the surface potential $\zeta$ and thus takes opposite signs at positively 
and negatively charged surfaces. All other contributions to $v_s$ are quadratic in $\zeta$. The self-propulsion 
velocity $u$ of a Janus particle is given by the surface average of the slip velocity, 
$\mathbf{u}=- \left<(1-\mathbf{n}\mathbf{n})\cdot\mathbf{v}_s\right>$, with the surface normal $\mathbf{n}$ \cite{And89}.

\begin{figure}[ptb]
\begin{center}
\includegraphics[width=9cm]{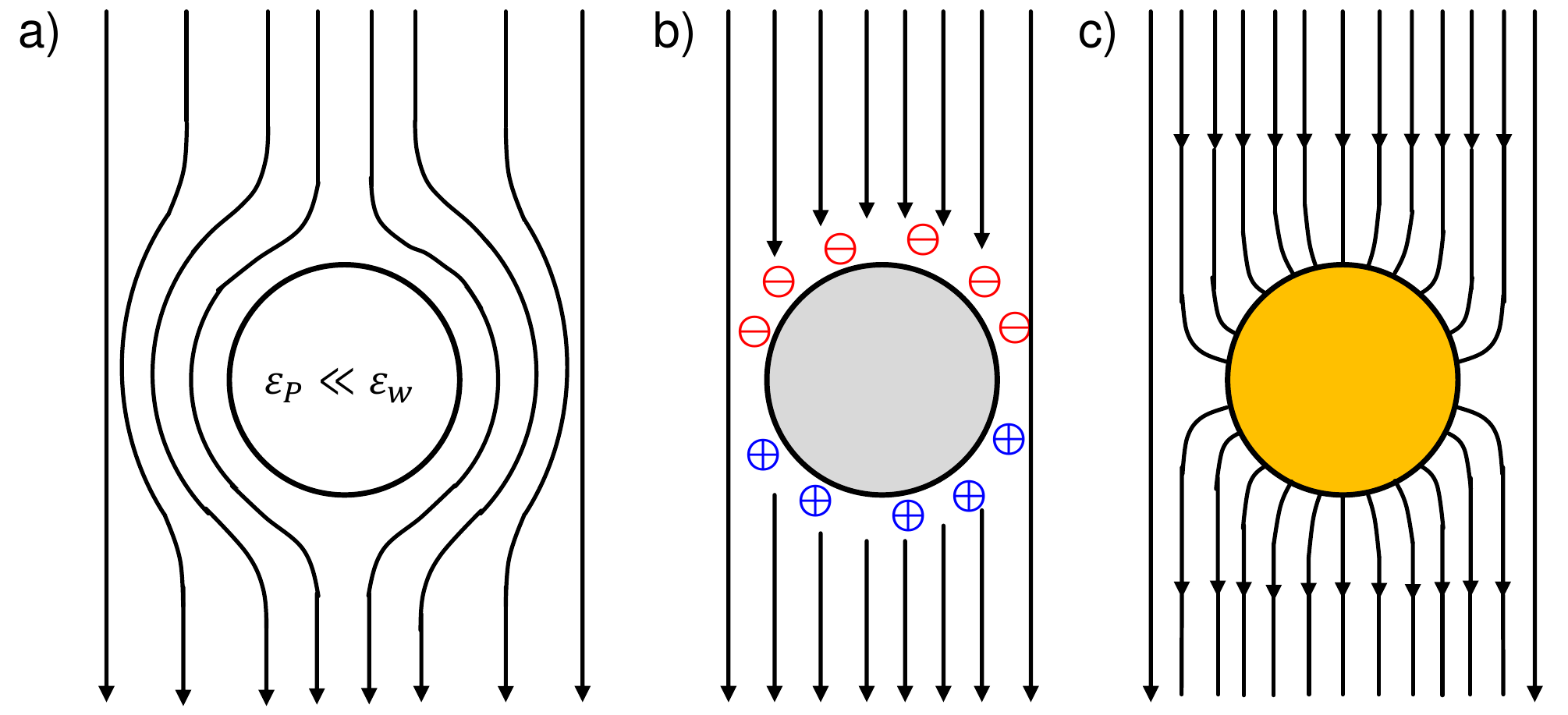}
\end{center}
\caption{Electric field lines for insulating and conducting particles. a) Electric field due to an applied 
external voltage. The field does not penetrate in a low-permittivity particle ($\varepsilon_P\ll \varepsilon_w$), 
resulting in a characteristic deformation. b) Thermoelectric field in the vicinity of an insulating particle. 
The field is not deformed by the permittivity contrast  but follows the temperature gradient, $E=S\nabla T$. 
(For the sake of simplicity we assume constant $\nabla T$, that is, similar thermal conductivities of particle 
and solvent.) Within one Debye layer from the particle surface, its normal component $E_\perp$ is screened by 
ion accumulation, that is, the thermocharge $\rho_T$, as shown in the left panel of Fig. 2; the parallel 
component $E_\parallel$ does not vanish, and the particle surface is not at constant potential. c) Thermoelectric 
field in the vicinity of a conducting particle. Polarization of the metal surface adjusts the surface charge 
density such that the parallel component of the field vanishes, resulting in an isopotential surface;  $\sigma$ is 
illustrated in the right panel of Fig.~2.  }
\label{fig:field}
\end{figure}

\section{Discussion}

Here we discuss the main features of the thermocharge and the thermoelectric field, and their dependence on material 
properties such as electric conductivity, surface roughness, and heat conductivity.

\subsection{Seebeck field in the vicinity of a spherical particle}

The Seebeck field does not result from an externally applied voltage but from thermocharges at the hot and 
cold boundaries which, in turn, are due to the thermal forces (\ref{eq0}) on the ions, as shown schematically 
in Fig. 1b for a Janus particle. At first sight one would expect that a thermoelectric field and an external 
field show the same behavior in the vicinity of a colloidal sphere. After all, both are subject to the same 
electrostatic boundary condition at the particle surface. It turns out, however, that their behavior is quite 
different. In Fig. \ref{fig:field} we compare their field lines around a spherical particle. For the sake of 
clarity we discuss the case of an external constant temperature gradient; the same physical effects occur for 
the self-generated gradient of a heated Janus particle or for a hot metal nanostructure.  

The left panel of Fig. \ref{fig:field}  shows the well-known deformation of an external electric field $E_0$ in 
the vicinity of a  low-permittivity particle. The parallel field at the surface varies as 
  \begin{equation}
  E_\parallel = \frac{3}{2}E_0\sin\theta
  \label{eq72}
  \end{equation}
with the polar angle $\theta$ \cite{And89}. With respect to the bulk field, it is enhanced by the permittivity 
ratio of particle and solvent, $3\varepsilon_w/(2\varepsilon_w + \varepsilon_P)\approx\frac{3}{2}$. 

The Seebeck field, on the contrary, results from surface charges; in 
order to satisfy the electric boundary condition for its normal component, it accumulates 
mobile ions with one screening length at the particle surface. The middle panel of Fig. \ref{fig:field} shows 
the thermoelectric field lines. They are not deformed and end at the thermocharge accumulated at the particle 
surface. The parallel component reads as
  \begin{equation}
  E_\parallel = S\nabla_\parallel T\sin\theta.
  \label{eq73}
  \end{equation}
Contrary to an external field, the surface field is not enhanced by the permittivity contrast.

The right panel shows the deformation of the Seebeck field by a conducting particle, where the parallel component 
of the surface field vanishes, $E_\parallel = 0$. From a comparison of the three situations shown, it is clear 
that the behavior of the thermoelectric field at solid boundaries significantly differs from that of a voltage induced field. 

The resulting electric field lines of a heated Janus particle are shown in Fig. 1b: The far-field corresponds to the 
Seebeck field (\ref{eq4}), whereas the near-field depends on the surface properties, as illustrated in Fig. 1c for 
the conducting and insulating hemispheres. The near-field corresponds to a superposition of the situations shown 
in Figs. \ref{fig:field} b and c.

\subsection{Thermocharge}

The thermocharge arises from the thermal forces $H_\pm \nabla T/T$ which drive the ions towards the hot or cold 
boundaries. When solving, in the simplest case, the zero-current condition (\ref{eq2}) and Gauss' law (\ref{eq3}), 
one finds that the steady state is characterized by a thermoelectric field and surface charges. The thermocharge per 
unit area $\sigma_T$, is independent of the material properties of the surface and of its surface charge $\sigma_0$. 
The profile of the diffuse layer, however, does depend on $\sigma_0$: At an uncharged surface, $\sigma_0=0$, it 
follows the exponential law $\rho_T=\sigma_Te^{-z/\lambda}$, whereas at a strongly charged surface, $\rho_T$ is 
part of the diffuse layer of Poisson-Boltzmann theory given in Eq. (\ref{B2}).

According to Eq. (\ref{eq8}), the thermocharge is entirely determined by the normal component of the temperature 
gradient at the solid surface and the Seebeck coefficient of the electrolyte, 
  \begin{equation} 
  \sigma_T = \varepsilon S\nabla_\perp T_S. 
  \end{equation}
On a sphere, the gradient is given by the local excess temperature and the radius, $\nabla_\perp T_S=-(T_S-T_0)/a$. In 
the case of a non-uniformly heated Janus particle, the temperature $T_S$ varies along the surface, and so does the 
charge per unit area $\sigma_T$, as illustrated in Fig. 1b.  A positive Seebeck coefficient, e.g., for aqueous solutions 
of NaCl, results in a negative $\sigma_T$, whereas a positive surface charge occurs for $S<0$ as, e.g., in NaOH solution.

As an estimate of its order of magnitude, we calculate the thermocharge density per unit area, $\sigma_T$, for a 
micron-size particle with an excess temperature of 30 K and the Seebeck coefficient of NaOH, $S=-200\,\mu\text{V/K}$, 
  \begin{equation}
   \sigma_T  \sim10^{-5}\,e/\mathrm{nm}^{2}. 
  \end{equation}
For comparison, the bare charge of a colloidal particle is of the order $\sigma_0\sim e/\mathrm{nm}^{2}$. 

\subsection{Polarization charge on a conducting surface}

The thermocharge discussed  above, is the same on insulating and conducting surfaces. On the latter, 
however, the isopotential condition of electrostatics imposes a polarization charge of the metal coating. Like 
any surface charge of a solid boundary, the polarization charge is screened by its  counterions. In other words, 
the polarization of the electronic system induces a corresponding polarization of the diffuse layer, as illustrated 
in right panel of Fig. 2. Thus the polarization effects concern only the immediate vicinity of the particle. Well 
beyond the Debye length, the effect of the polarization charges vanishes. Accordingly, the field lines of insulating 
and conducting particles in Fig. \ref{fig:field}b and c, differ within the screening length, but are identical at larger distances.  

For an excess temperature of 30 K, the Seebeck coefficient $S=-200\,\mu\text{V/K}$, and $\lambda=2$ nm, the 
weak-coupling expression (\ref{eq35}) gives the order-of-magnitude estimate
  \begin{equation}
   \sigma_P  \sim10^{-2} e/\mathrm{nm}^2.
  \end{equation}
When comparing with the thermocharge, one finds that  $\sigma_P$ exceeds $\sigma_T$ by a factor $a/\lambda$ which, for 
micron-size particles, is of the order of $a/\lambda\sim1000$. On the other hand, the polarization charge may attain 
several percent of the colloidal surface charge $\sigma_0$. 

As a related quantity we estimate the thermopotential $\varphi_T=-S(T_S -T_0)$. The above parameters 
give  $\varphi_T\sim6$ mV, which is almost comparable to the surface potential of moderately charged 
colloids, $\zeta\sim30$ mV. One should note, however, that the variation of $\varphi_T$ is  a more 
relevant quantity than its absolute value. Still, for a typical temperature profile, one finds the 
thermopotential at the two poles of a Janus particle differs by about half of its mean value.  

\begin{figure}[ptb]
\begin{center}
\includegraphics[width=12cm]{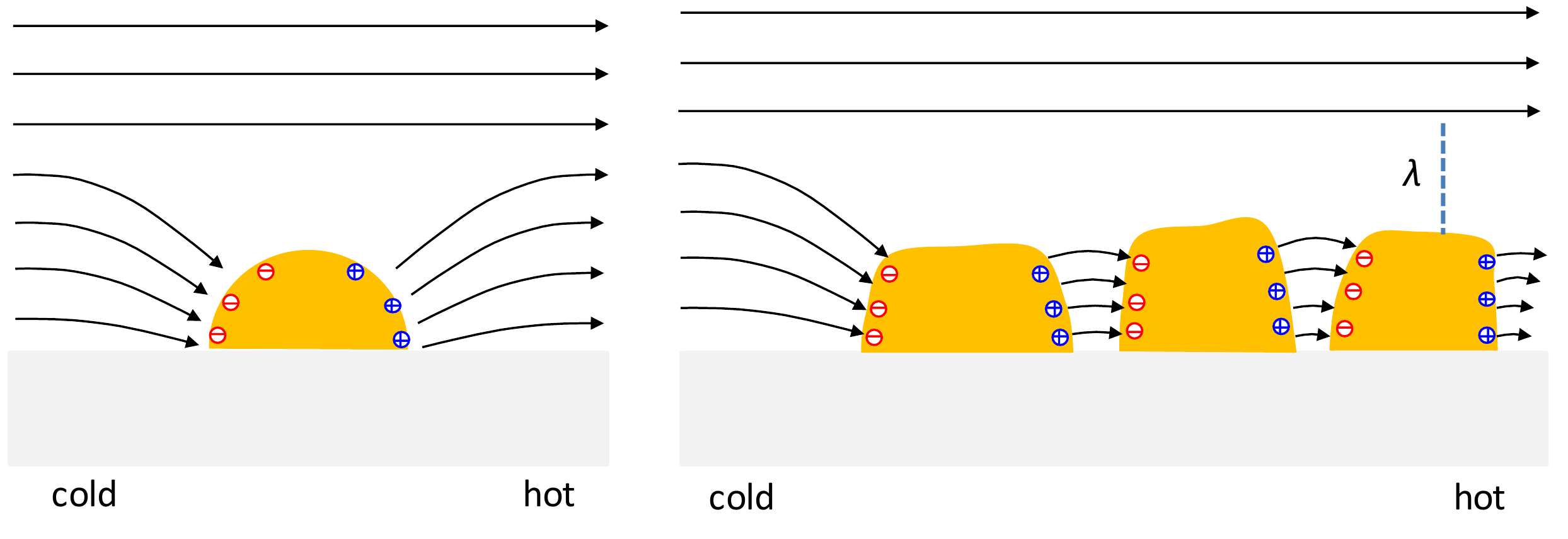}
\end{center}
\caption{Schematic view of the thermoelectric field and polarization charges of gold grains at a low-permittivity 
insulating surface, in contact with an electrolyte solution with positive Seebeck coefficient, $S>0$. The left 
panel shows the case of single grain. The field is normal at the grain surface because of the electrostatic 
boundary conditions, and parallel at the insulating solid, because of the strong permittivity contrast of its 
material and water. Well beyond the screening length one recovers the constant Seebeck far-field. The right panel 
shows a densely covered surface, where each grain forms an equipotential surface and carries opposite polarization 
charges $\protect\sigma _{P}$ at its cold and hot sides.}
\label{Fig:grains}
\end{figure}

\subsection{Granular gold surface} 

So far we considered a continuous gold surface, as shown in Fig. 1c. 
Yet this does not always correspond to the actual experimental
situation. For example, the cap of the Janus particles used in Ref.\ \cite{Sim16}
consists of a dense coverage of nano-sized gold grains, visible in scanning
electron microscopy images \cite{Ned15}. Since the grains are not connected, 
the active cap of these particles does not form an isopotential surface, contrary 
to what we assumed so far. 

Here we give a qualitative discussion of the resulting Seebeck field and slip velocity. 
From our results for conducting surfaces, it is clear that the parallel component of the 
thermoelectric field is screened within one screening length. 
Fig.~\ref{Fig:grains} gives a schematic view of an insulating surface, partly covered 
by gold grains and at a non-uniform temperature. According to the discussion in Sect. \ref{sect:temp} 
below, we neglect the thermal conductivity contrast. On the other hand, gold nanostructures keep their 
electric conductivity, though it is lower than that of bulk material; thus the grains are conducting and that 
each of them forms an equipotential surface. 

The left panel of Fig. \ref{Fig:grains} shows the thermoelectric properties of a single grain. The 
parallel component of the Seebeck far-field induces polarization charges, such that isopotential condition 
is satisfied at the surface. For a gold hemisphere 
mounted on a low-permittiivity material, the resulting electric field and polarization charges are obtained from 
Appendix A.3 by retaining the term $c_1$ only. Note that beyond a distance of one screening length, 
one recovers the macroscopic thermoelectric field. 
The right panel of Fig. \ref{Fig:grains} shows a densely covered surface, where the distance between grains does not 
exceed the screening length. Then the overall
thermopotential is split in small jumps between nearby grains; their cold and hot boundaries carry 
polarization charges which result in a strong electric field in the spacing. The field component parallel to 
the surface vanishes at the grain surface but increases beyond and tends towards the far-field value beyond 
double layer. 

The slip velocity is essentially determined by the layer of thickness $\lambda$ above the gold grains, whereas 
the narrow space between the grains is of little relevance. A different picture would arise if the gold grains covered 
only a small fraction of the surface, and if their height was small as compared to their spacing. For a situation as 
shown by Fig. \ref{Fig:grains} or by the electron micrograph in Ref. \cite{Ned15}, however, we conclude that the 
picture developed for micron-size conducting surfaces remains at least qualitatively correct for a granular gold coating.  
Because of the surface roughness one may expect a somewhat modified slip velocity, probably smaller than at a homogeneous cap.

\subsection{Comparison with experiment}

So far there are few direct measurements of the slip velocity with respect to a wall \cite{Der41,Bre16}; most 
experiments report the motion of dispersed particles in an external temperature 
gradient \cite{Bra13,Lin17,Lin17b,Put05,Vig10,Esl14} or of self-propelling microswimmers \cite{Jia10,Bar13,Sim16,But12}, 
where the velocity is given by the surface average of the slip velocity.

The slip velocity (\ref{eq48}) consists of various contributions, proportional to the temperature gradient and 
its companion fields. All of them are of comparable magnitude. The slip velocity varies as a function of the 
electrolyte strength and, through the Soret and Seebeck coefficients, depends on specific-ion properties. At 
room temperature, the Soret coefficient $S_T$ is usually positive \cite{Nak88}. Then except for the Seebeck 
contribution, all terms of the slip velocity are positive, and the boundary layer flows towards the hot. There 
is, however, strong evidence that the Seebeck term is dominant for common salt and buffer solutions, such as 
NaCl, NaOH, citric acid, and CAPS. Since their Seebeck coefficients take opposite signs, one observes, as a most 
striking feature, a positive slip velocity for NaCl \cite{Esl14} and a negative one for NaOH \cite{Vig10,Esl14}; 
thus changing the anion reverses the direction of thermally driven motion. Similar effects were reported for 
buffer solutions \cite{Put05}. 

The main results of the present paper concern the slip velocity along metal surfaces. Though local heating of 
gold structures is widely used  for manipulating of particles and cells \cite{Bra13,Lin17,Lin17b,Lin17a} or 
powering microswimmers \cite{Jia10,Bar13,Sim16,But12,Gir16}, there is at present no systematic study of the 
creep flow along a conducting surface. Evidence for thermo-electric driving of hot silica particles with a 
granular gold cap, was reported by one recent experiment  \cite{Sim16}: Probing the particle's self-propulsion 
velocity in 10 mM solutions of NaCl, LiCl and NaOH, revealed
a salt-specific effect, which agrees qualitatively with the Seebeck coefficients of these 
electrolytes, $S_{\text{NaCl}}>S_{\text{LiCl}}>S_{\text{NaOH}}$. Since the thermophoretic 
self-propulsion is superposed on motion due to radiation pressure and gravity, these data do not 
provide an absolute measure of $v_s$, but only qualitative differences upon changing the 
ions. In summary, the data of Ref. \cite{Sim16} confirm the existence of an electrolyte 
Seebeck effect for active Janus particles, yet they do not provide clear evidence whether 
the thermoelectric driving is the same on the silica and gold hemispheres, as suggested by 
the present work, or whether the Seebeck effect vanishes on the metal surface.

\subsection{Temperature gradient at the particle surface}\label{sect:temp}

Throughout this paper we have assumed that the temperature gradient is not modified at the solid-water 
interface, which is justified as long as the heat conductivities of liquid and particle, $\kappa_w$ and $\kappa_P$, 
take similar values. For a sufficiently strong conductivity contrast, however, the particle deforms the temperature 
field in its vicinity. For a sphere, a conductivity contrast modifies the parallel and perpendicular components 
of the temperature gradient according to 
  \begin{equation}
  \nabla_\parallel  T_S\rightarrow  \xi_\parallel\nabla_\parallel  T_S, \;\;\;\;\; 
                 \nabla_\perp  T_S\rightarrow  \xi_\perp\nabla_\perp  T_S,
   \label{eq74}
   \end{equation}
with the well-known constants \cite{Wue10}
  \begin{equation}
        \xi_\parallel = \frac{3\kappa_w}{2\kappa_w + \kappa_P}, \;\;\;
                        \xi_\perp = \frac{3\kappa_P}{2\kappa_w + \kappa_P}.
   \label{}
   \end{equation}
In order to account for the conductivity contrast in the results of the preceding sections, one merely 
has to introduce these factors. Typical insulating materials, like silica and polystyrene, show a somewhat 
lower heat conductivity than water. The most important thermoelectric properties are proportional to 
parallel gradient, with a correction factor  $\xi_\parallel $ between 1 and $\frac{3}{2}$, which is 
usually of little relevance. 

A more complex situation occurs for thin metal coatings. Metals conduct heat much better 
than water, $\kappa_m\gg \kappa_w$. A metal coating of thickness $d$ significantly deforms 
the temperature profile of a sphere of radius $a$, if the conductivity contrast 
satisfies $\kappa_m/\kappa_w>a/d$; in the thick-cap limit the metal surface becomes an 
isothermal \cite{Bic13}. Noting that $\kappa_m$ decreases for films of less than $100\,\mathrm{nm}$, one 
finds that for micron-size particles, the temperature is modified by coatings thicker than 
several tens of nanometer. 

Most recent experiments are done on Janus particles with thinner coatings, of less than $10\,\mathrm{nm}$, 
where the cap contribution to heat conduction and the resulting deformation of the temperature field can 
be neglected. On the other hand, such thin gold coatings still have significant electrical conductivity, 
and thus develop polarization charges as discussed in this paper and shown in Figs. 2 and 4.   

\subsection{Transient and memory effects.}

So far we have considered the
steady-state Seebeck effect. The transient behavior after switching on the
heat source is readily obtained from the advection-diffusion equation for
the ions with Gauss' law $\nabla \cdot E=\rho /\varepsilon $. Thus we find 
\begin{equation}
\rho _{T}(t)=\rho _{T}(\infty )(1-e^{-t/\tau_\text{ion} }),
\end{equation}
where the characteristic time scale expresses the time of ion diffusion over
the screening length, 
\begin{equation}
\tau _{\text{ion}}=\frac{\lambda ^{2}}{2D}.
\end{equation}%
With typical values $D\sim 10^{-9}$ m$^{2}$/s and $\lambda \sim 50$ nm, one
finds $\tau \sim \mu $s. 

Thus building up the Seebeck field requires a few microseconds, 
and the same time-dependence occurs for the slip velocity.
Indeed, the thermal and hydrodynamic time scales, $\tau _{\text{th}}=\lambda
^{2}/\alpha $ and $\tau _{\text{hy}}=\lambda ^{2}/\nu $, are by several
orders of magnitude shorter, since both heat diffusivity ($\alpha \sim
10^{-7}$ m$^{2}$/s) and kinematic viscosity ($\nu \sim 10^{-6}$ m$^{2}$/s),
by far exceed ion diffusivity. It should be noted that $\tau _{\text{hy}}$
is much shorter than the hydrodynamic memory of Brownian motion, $\tau _{%
\text{hy}}^{\prime }=a^{2}/\nu $ \cite{Fra11}; this is due to the fact that
in the latter case, the hydrodynamic stress decays over the particle size $a$%
, whereas for phoretic and active particles, the relevant stress is confined
within the interaction length $\lambda $ \cite{Fay08}.

As a consequence, we expect a rather intricate behavior of the particle
motion during the first milliseconds, 
  \begin{equation}
  v_{s}(t)=v_{s}^{\text{el}}(1-e^{-t/\tau _{\text{ion}}})  
            +  v_{s}^{\text{osm}}(1-e^{-t/\tau _{\text{th}}}),
  \label{eq60}
  \end{equation}
where the thermo-electric slip velocity $v_{s}^{\text{el}}$ corresponds to the first 
term of (\ref{eq48}), and the osmosis-driven one $v_{s}^{\text{osm}}$ to the remainder. 
The latter sets in on the heat-diffusion time scale $\tau _{\text{th}}\sim 10$ nanoseconds. 
The Seebeck effect requires ion diffusion which occurs on the time scale $\tau _{%
\text{ion}}$ that may attain a microsecond. Since in many instances, the
thermoelectric slip velocity $v_{s}^{\text{el}}$ is stronger and carries the
opposite sign \cite{Put05,Vig10,Esl14}, the onset of the Seebeck effect could even
result in a reversal of the direction of motion.

The above discussion applies to the double-layer at the conducting
hemisphere, where the local temperature gradient is determined by absorption
of laser light by the gold coating. At the insulating hemisphere, 
building up the stationary temperature profile requires heat diffusion over
a distance comparable to the particle radius. Thus the thermal time scale, 
$\tau _{\text{th}}^{\text{ins}}=a^{2}/\alpha $, is of the order of ten
microseconds, which is close to the ionic relaxation time $\tau_\text{ion}$. 
Thus on an insulating surface, the time scales of the two terms in the slip velocity 
(\ref{eq60}) are not very different. 
 
\section{Summary}

We find that hot metal structure in contact with an electrolyte solution, show thermoelectric 
properties at the nanoscale that depend both on  surface material properties and ion-specific effects. 
Here we briefly summarize our main results. 

The diffuse layer comprises a thermocharge $\rho_T$ which is proportional to the 
surface temperature $T_S$. On a Janus particle, $T_S$ increases from 
the passive hemisphere to the heated cap, and so does the thermocharge, resulting 
in a parallel component of the Seebeck field along the particle surface.

On a conducting surface, such as a gold cap, however, the parallel temperature
gradient induces a polarization charge on the metal structure, which modifies the 
double layer such that the parallel component of the electric field vanishes at the 
surface. Yet this does not affect the thermally induced slip velocity, which 
turns out to be identical on insulating and conducting surfaces. 

In previous work the Seebeck field had been considered like a field due to an external voltage. 
We find that the near-field is rather different, as shown in Figs. 3 a and b. As a consequence, the 
parallel field at a spherical particle (\ref{eq73}) does not carry a factor $\frac{3}{2}$, contrary to 
an external field (\ref{eq72}) at a low-permittivity particles. The same difference occurs between the 
thermoelectric contribution to the slip velocity (\ref{eq48}) and the electroosmotic velocity. 

Regarding specific-ion effects, our findings agree qualitatively with a recent
experiment on gold-capped silica particles,  showing a significant variation of the self-propulsion 
velocity with the used salts NaOH, NaCl, LiCl \cite{Sim16}. The data do not provide 
conclusive evidence for thermoelectric driving along the metal cap. 

From our analysis of the onset of thermoosmotic and thermoelectric driving, we expect 
striking transient effects. Because of the slow diffusion of ions, as compared to
diffusion of heat and momentum, the thermo-electric slip velocity sets in on
a microsecond timescale. The much faster onset of thermoosmosis, should result in a 
two-step transient behavior upon switching on the heating.

AL and AW acknowledge support by the French National reasearch agency through
contract ANR-13-IS04-0003. AW thanks Frank Cichos and Martin Fr\"anzl for stimulating 
discussions.

\appendix

\section{Thermocharge of an uncharged particle}

Here we derive in detail the thermocharge of an uncharged spherical particle. Since the 
thermocharge is small, we may resort to Debye-H\"uckel approximation, and for the 
spherical geometry, the thermoelectric potential can be given explicitly in terms of 
a multipole expansion. We first provide the general formulae for weak coupling, evaluate 
them for a 1D geometry, and then consider insulating and conducting particles with 
non-uniform surface temperature.

\subsection{Debye-H\"uckel theory}

Here we derive the thermocharge of an uncharged hot particle in some more detail 
than in the main text. We start from the relation between thermocharge density $\rho$ 
and thermoelectric field $ \mathbf{E}$ obtained in Sect. II,
\begin{equation}
         \nabla \rho + \frac{\varepsilon}{\lambda^2}  ( S\mathbf{\nabla }T - \mathbf{E} )= 0 .
\label{A2}
\end{equation}
This equation has two solutions, and the electrostatic potential consists of two contributions 
accordingly, 
\begin{equation}
\varphi =\varphi _{T}+\varphi_\sigma.  
\label{A4}
\end{equation}%
The first one, $\mathbf{E}_T=S\mathbf{\nabla }T$ and $\rho=0$, corresponds to the far-field 
(\ref{eq4}) with zero charge density and the thermoelectric potential 
  \begin{equation}
  \varphi_T=-S(T-T_{0}),  
   \label{A5}
   \end{equation}
whereas the second solution is given by the screened Debye-H\"uckel potential  $\varphi_\sigma$. 
Indeed, completing $\nabla \rho = (\varepsilon/\lambda^2) \mathbf{E}$ with Gauss' law 
$\rho = \varepsilon  \text{div} \mathbf{E}$, one finds  
$\mathbf{E}=- \mathbf{\nabla}\varphi_ {\sigma}$, where $\varphi_{\sigma}$ solves the 
Debye-H\"uckel equation 
   \begin{equation}
   \nabla^2\varphi_\sigma=\frac{\varphi_\sigma}{\lambda^2}. 
   \end{equation}

\subsection{1D geometry}

These equations have been solved previously for a 1D geometry between a hot and a 
cold plate \cite{Maj11}, and for a uniformly heated spherical particle \cite{Maj12}. With 
the constant temperature gradient $\nabla T$ along the $z-$direction, one readily 
calculates the Seebeck field,
   \begin{equation}
   E = S\nabla T  \left( 1 - \frac{\cosh(z/\lambda)}{\cosh(L/2\lambda)} \right),
   \end{equation}
where $-L/2\le z\le L/2$. If the system size $L$ is much larger than the Debye 
length $\lambda$, one has the bulk field $E = S\nabla T$; at the boundaries, $E$ is 
exponentially screened and vanishes at the hot and cold surfaces. 

The corresponding thermocharge at the boundaries is given by Gauss' law, 
   \begin{equation}
   \rho_T = \varepsilon \partial_z E =  - \varepsilon S\nabla T\frac{\sinh(z/\lambda)}{\cosh(L/2\lambda)}.
   \end{equation}
For $L\gg \lambda$, this simplifies to $\rho_T =\mp \varepsilon S\nabla T e^{(\pm z-L/2)/\lambda}$, resulting 
in positive and negative charge layers at the hot and cold boundaries.  

\subsection{Spherical particle}

For a spherical particle, the inhomogeneous solution (\ref{A5}) is given by a multipole expansion for
the temperature field, 
\begin{equation}
T(\mathbf{r})=T_{0}+\sum_{n=0}^{\infty }t_{n}P_{n}\left( c\right) \frac{%
a^{n+1}}{r^{n+1}},  \label{eq19}
\end{equation}%
where $c=\cos \theta $ is the cosine of the polar angle. The mean excess surface
temperature $t_{0}=q/4\pi \kappa a$ is determined by the rate of heat
absorption $q$, the thermal conductivity of the solvent $\kappa $, and the
particle radius $a$.

The homogeneous solution $\varphi_\sigma$ is obtained as a series  
  \begin{equation}
      \varphi _\sigma=\sum_{n=0}^{\infty }c_{n}P_{n}
                       \left( c\right) \frac{k_{n}\left(r/\lambda \right) }{k_{n}\left( a/\lambda \right) },  
  \label{14}
  \end{equation}
in terms of Legendre polynomials $P_{n}\left( c\right) $ with $c=\cos \theta 
$, and the modified spherical Bessel function of the second kind $%
k_{n}\left( x\right) $. For the sake of notational convenience, we introduce
the factor $k_{n}\left( a/\lambda \right) $, such that the radial solutions
are normalized at the particle surface $r=a$. 

\subsection{Insulating particle}

The coefficients $c_{n}$ of the homogeneous solution remain to be determined from the 
electrostatic boundary conditions at the particle-water interface. For a low-permittivity 
material we may put $\varepsilon_P/\varepsilon_w\rightarrow 0$. Then the boundary
conditions require that the normal component of the electric field vanishes, 
  \begin{equation}
  E_{\bot}(r=a)=0.
  \label{A10}
  \end{equation}
Taking the radial derivative of $\varphi $, putting $r=a$, and rearranging
terms we find 
  \begin{equation*}
   c_{n}=St_{n}(n+1)\frac{\lambda }{a}\frac{k_{n}\left( a/\lambda \right) }
                                                         {k_{n}^{\prime }\left( a/\lambda \right) }.
  \end{equation*}%
with the dimensionless derivative $k_{n}^{\prime }\left( x\right) =\partial_{x}k_{n}(x)$.

These coefficients determine the thermopotential $\varphi$. In order to simplify the resulting 
expressions we note that the ratio of Debye length $\lambda $ and particle radius $a$ is at 
most of the order of a few percent. Expanding in powers of the small parameter $%
\lambda /a$, 
\begin{equation*}
\frac{k_{n}\left( r/\lambda \right) }{k_{n}\left( a/\lambda \right) }=\frac{a%
}{r}e^{\left( a-r\right) /\lambda }\left[ 1+\frac{\lambda }{a}\frac{n\left(
n+1\right) }{2}\left( 1-\frac{a}{r}\right) +...\right] ,
\end{equation*}%
we find that the first terms of the series are well approximated by 
\begin{equation*}
\frac{k_{n}\left( r/\lambda \right) }{k_{n}\left( a/\lambda \right) }\approx 
\frac{a}{r}e^{\left( a-r\right) /\lambda }\ \ \ \ (n<\sqrt{a/\lambda }).
\end{equation*}%
In the most relevant near-field range, this approximation is even valid for $%
n<a/\lambda $. To leading order in the small parameter $\lambda /a$, we have 
$k_{n}^{\prime }\left( a/\lambda \right) /k_{n}\left( a/\lambda \right)
= - 1+O(\lambda /a)$. Then the above coefficient simplifies according to 
  \begin{equation}
  c_{n} = St_{n}(n+1)\frac{\lambda }{a},
  \end{equation}%
and the electrostatic potential reads as 
  \begin{equation}
    \varphi  = -S\sum_{n}t_{n}P_{n}(c)\left( \frac{a^{n+1}}{r^{n+1}}
                           -(n+1)\frac{\lambda }{r}e^{\left( a-r\right) /\lambda }\right) .
  \label{A12}
\end{equation}%
The screened term is by a factor $\lambda /a$ smaller than the first one;
yet their radial derivatives cancel each other at $r=a$, thus satisfying (\ref{A10}).

The normal component of the electric field reads, to leading order in $\lambda/a$,
  \begin{equation}
    E_\perp(\mathbf{r})  = S\nabla_\perp T(\mathbf{r}) (1 - e^{(a-r)/\lambda }) .
  \label{A14a}
\end{equation}
In the screened terms we have discarded factors of $a/r$, since they are close to
unity in the range where the exponential function is finite. This explicits how the 
thermocharge screens the normal electric field. The parallel field component, on the 
contrary, is hardly affected by the thermocharge,
  \begin{equation}
    E_\parallel(\mathbf{r})  = S\nabla_\parallel T(\mathbf{r}) (1 + O(\lambda/a)) .
  \label{A15}
\end{equation}

The thermocharge density follows from Gauss' law, $\rho_T=-\nabla^2\varphi_\sigma$. 
With the same approximations as for the normal field component above, we  have 
  \begin{equation}
  \rho _{T} = \frac{\varepsilon }{\lambda }e^{\left( a-r\right) /\lambda}S\nabla _{\bot }T|_{S}. 
   \label{42} 
\end{equation}%
Integrating over the radial coordinate we find the
charge per unit area 
  \begin{equation}
  \sigma_T = \int_{0}^{\infty }dr\rho _{T} = \varepsilon S\nabla _{\bot }T|_{S}.
  \label{A16}
  \end{equation}%
Integrating over the particle surface gives the total charge
  \begin{equation}
      Q_{T}=-4\pi a\varepsilon St_{0},
   \end{equation}%
which is determined by the isotropic component of the excess temperature.

\subsection{Conducting particle}

The thermocharge $\rho_T$ is the same as obtained above for an insulating particle. 
The  boundary conditions, however, impose that the parallel component of the electric field vanishes, 
whereas the normal component is eventually compensated by a polarization charge density $\sigma_P$ 
of the surface $r=a$,
  \begin{equation}
    E_\perp  =  -\frac{\sigma_P}{\varepsilon},\;\;\;\;   E_\parallel  = 0  .
  \label{A18}
\end{equation}
Writing the surface charge as a series  $\sigma_P=\sum_n s_n P_n(c)$ and inserting the 
potential $\varphi$, we determine the coefficients $c_n$ 
and $s_n$ to leading order in $\lambda/a$,
  \begin{equation*}
   c_{n}=  St_{n},  \;\;\;\;  s_n =  -  \frac{\varepsilon St_n}{\lambda} \;\;\;\; \;\; (n>0).
  \end{equation*} 
The isotropic terms are particular because of charge conservation,  
  \begin{equation*}
   c_{0}=   \frac{\lambda}{a}St_{0}, \;\;\;\;  s_0 =  0 .
  \end{equation*}%
Then the electrostatic potential reads
  \begin{equation}
    \varphi  =      -St_{0} \frac{a-\lambda e^{\frac{a-r}{\lambda }}}{r} 
                               -S\sum_{n>0}t_{n}P_{n}(c)\left( \frac{a^{n+1}}{r^{n+1}}
                                             - e^{\frac{a-r}{\lambda }}\right) .
  \label{A12a}
\end{equation}%
Resorting to the same approximation as in the insulating case, we have
  \begin{equation}
    \varphi  = -St_{0} \frac{a-\lambda e^{\frac{a-r}{\lambda }}}{r}
                          -S ( T_S-\left<T_S\right>) \left(1 - e^{\frac{a-r}{\lambda }}\right) ,
  \label{A13}
\end{equation}
with the surface temperature $T_S$ and its mean value $\left<T_S\right>$. 
The polarization charge is given by 
  \begin{equation}
    \sigma_P  = \frac{\varepsilon}{\lambda} S (T_S-\left<T_S\right>) , 
  \label{A14}
\end{equation}

For the normal component of the electric field we find
  \begin{equation}
    E_\perp(\mathbf{r})  = S\nabla_\perp T(\mathbf{r})  
                                      - \frac{S (T_S-\left<T_S\right>)}{\lambda}e^{\frac{a-r}{\lambda }} .
  \label{A16a}
\end{equation}
In the screened terms we have discarded factors of $a/r$, since they are close to
unity in the range where the exponential function is finite. This explicits how the 
thermocharge screens the normal electric field. 

To linear order in the excess temperature, the parallel field component,
  \begin{equation}
    E_\parallel(\mathbf{r})  = S\nabla_\parallel T \left(1 -  e^{\left( a-r\right) /\lambda }\right), 
  \label{A18a}
\end{equation}
vanishes at the surface and tends to the Seebeck field well beyond the double layer. 

The thermocharge density follows from Gauss' law, $\rho_T=-\nabla^2\varphi_\sigma$. 
With the same approximations as for the normal field component above, we  have 
  \begin{equation}
  \rho _{P} = \frac{\varepsilon }{\lambda^2 }e^{\left( a-r\right) /\lambda}S (T_S-\left<T_S\right>).  
\label{A20}
\end{equation}%

\section{Poisson-Boltzmann theory}

Consider a charged surface in contact with an electrolyte solution. In mean-field theory, 
the electrostatic potential $\varphi_\sigma$ satisfies the Poisson-Boltzmann equation 
  \begin{equation}
  \nabla ^{2}\varphi_\sigma = -\frac{\rho}{\varepsilon} 
                                             = \frac{k_{B}T}{e\lambda ^{2}}\sinh \frac{e\varphi_\sigma}{k_{B}T}.  
  \label{B2}
  \end{equation}%
If the particle radius is much larger than the Debye screening length, the
curvature of the surface can be neglected. Then the Laplace operator reduces
to the second derivative with respect to the vertical coordinate $z$, and
the potential is the 1D solution \cite{And06} 
  \begin{equation}
  \varphi_\sigma(z) = - \frac{2k_{B}T}{e}\ln \frac{1+ge^{-z/\lambda }}{%
  1-ge^{-z/\lambda }},  
  \label{B3}
  \end{equation}%
with the shorthand notation 
  \begin{equation} 
  \hat{g}=ge^{-z/\lambda }, \;\;\;\;  g=\sqrt{1+\ell ^{2}/\lambda ^{2}}-\ell /\lambda.
  \label{B3a}
  \end{equation}%
The parameter $g$ is given by the ratio of the Gouy-Chapman length $\ell=e/(2\pi \ell _{B}|\sigma|)$ and 
the Debye length $\lambda =1/\sqrt{8\pi \ell _{B}n}$. With the Bjerrum length $\ell _{B}=e^2/(4\pi \varepsilon k_BT)$ one 
finds 
  \begin{equation}
  \frac{\ell }{\lambda} = \frac{e}{2\pi |\sigma| \ell_B \lambda} = \frac{e \sqrt{8n \varepsilon k_BT}}{|\sigma|} ,
  \label{B4}
  \end{equation}
with the salinity $n$. In the following we assume that $\sigma$ is positive, corresponding 
to the usual situation of a negative surface charge $-\sigma$. 

The electric field $E_\perp=-d\varphi/dz$ is perpendicular to the surface and reads 
  \begin{equation}
   E_\perp  = -\frac{\sigma}{\varepsilon }e^{-z/\lambda }\frac{1-g^{2}}{1-\hat{g}^{2}}.
  \label{B5}
  \end{equation}
At the particle surface, one readily verifies the relation $E(0)=-\sigma/\varepsilon $.

The charge density $\rho$ in the diffuse layer is given by the second equality in (\ref{B2}). An 
equivalent form in terms of the parameter $g$ is obtained from Gauss' law $\rho=\varepsilon dE/dz$, 
  \begin{equation}
  \rho  = \frac{\sigma}{\lambda}e^{-z/\lambda }
                                                             \frac{(1-g^2)(1+\hat{g}^{2})}{1-\hat{g}^{2}}.  
  \label{B6}
  \end{equation}
Integrating over $z$ one finds
  \begin{equation}
     \int_0^\infty dz \rho(z)  = \sigma,  
  \label{B7}
  \end{equation}
which is opposite to the surface charge density $-\sigma$. We also give the excess ion concentration 
  \begin{equation}
  \delta n = n_+ + n_- - 2n = 2n \left( \cosh\frac{e\varphi_\sigma}{k_B T} -1 \right).  
  \label{B8}
  \end{equation}

The Debye-H\"uckel approximation is obtained by taking the limit of small surface charge, where 
$\ell/\lambda\gg1$ and $g= \frac{1}{2}\lambda/\ell$, resulting in
$$  
    \varphi_\sigma  = - \frac{\sigma\lambda}{\varepsilon}e^{-z/\lambda },\;\;\;\;
    E_\perp  = - \frac{\sigma}{\varepsilon}e^{-z/\lambda },\;\;\;\;
    \rho  =  \frac{\sigma}{\lambda}e^{-z/\lambda },\;\;\;\; 
    \delta n = n \left( \frac{e\varphi_\sigma}{k_B T} \right)^2.  
$$

\section{Determination of the polarization charge $\sigma_P$}

Anticipating that the $\sigma_p$ is much smaller than the uniform surface charge $\sigma_0$,
we expand the Poisson-Boltzmann potential to linear order, 
  \begin{equation}
  \varphi_\sigma = \varphi_{\sigma_0} + \sigma_P\frac{d\varphi_{\sigma_0}}{d\sigma_0}.
  \label{A30}
  \end{equation}
Taking the parallel gradient component, we have 
\begin{equation}
 \nabla _{\Vert }\varphi_\sigma = \nabla _{\Vert }\varphi _{\sigma _{0}} -\frac{\nabla _{\Vert
}\sigma _{P}}{\sigma _{0}}\frac{2k_{B}T}{e \sqrt{1+b^{2}}},
\end{equation}
where $b=\ell/\lambda$ is the ratio of the Gouy-Chapman length $\ell$ and the Debye length $\lambda$.
Noting that this gradient vanishes at the surface ($z=0$) and solving for $\nabla _{\Vert }\sigma _{P}$, 
we obtain
\begin{equation}
\frac{\nabla _{\Vert }\sigma _{P}}{\sigma _{0}}=-\frac{e}{2k_{B}T}\sqrt{%
1+b^{2}}\left( S\nabla _{\Vert }T-\nabla _{\Vert }\varphi _{\sigma_0}\right) .
\label{86}
\end{equation}

Now we compute the last term in parentheses at $z=0$
  \begin{equation}
    \nabla _{\Vert }\varphi _{\sigma _0} =\zeta \frac{\nabla
       _{\Vert }T}{T}+\frac{k_{B}T}{e}\frac{1}{\sqrt{1+b^{2}}}\left( \frac{\nabla
            _{\Vert }\varepsilon }{\varepsilon }+\frac{\nabla _{\Vert }T}{T}\right) .
  \label{87}
  \end{equation}
Inserting this in Eq. (\ref{86}), we obtain finally the surface
charge $\sigma _{P}$ in Poisson-Boltzmann theory as,%
    \begin{eqnarray}
    \frac{\nabla _\Vert \sigma _P}{\sigma_0} =-\frac{e\sqrt{1+b^2}}{2k_{B}T}
     \left( S\nabla _{\Vert }T-\zeta \frac{\nabla _{\Vert }T}{T}\right)
         +\frac{1}{2}\left( \frac{\nabla
_{\Vert }\varepsilon }{\varepsilon }+\frac{\nabla _{\Vert }T}{T}\right).
\end{eqnarray}
With the permittivity gradient $\nabla\varepsilon=(d\varepsilon/dT)\nabla T$, 
we obtain  the integral 
    \begin{equation}
    \frac{\sigma _P}{\sigma_0} = - \left(\frac{e\sqrt{1+b^2}}{k_{B}T}
     \left( ST-\zeta \right) - \frac{d\ln\varepsilon}{d\ln T} - 1 \right)
                   \frac{\Delta T - \left<\Delta T\right>}{2T} .
\end{equation}
The last factor follows from the condition of charge neutrality, 
$$\left<\sigma_P\right> = \frac{1}{S}\int_{S}\sigma _{P}dS=0.$$

In the weak-coupling limit,  the Gouy- Chapman length is large as compared to the 
Debye length, $b\gg 1$. Expanding in first order in $b^{-1}$, we find
the surface polarization charge in Debye- H\"{u}ckel approximation as,%
\begin{equation}
\sigma _{P}=\frac{\varepsilon S}{\lambda }\left( \Delta T-\left\langle \Delta
T\right\rangle \right),   \;\;\;\;\;\;\;\; (\ell/\lambda \ll 1) .
\label{eq92}
\end{equation}

\end{document}